\documentclass{aa}  
\usepackage{graphicx}
\usepackage{txfonts}
\usepackage{natbib,twoopt}
\usepackage[breaklinks=true]{hyperref} 
\bibpunct{(}{)}{;}{a}{}{,}             
\makeatletter
  \newcommandtwoopt{\citeads}[3][][]{\href{http://adsabs.harvard.edu/abs/#3}%
    {\def\hyper@linkstart##1##2{}%
     \let\hyper@linkend\@empty\citealp[#1][#2]{#3}}}
  \newcommandtwoopt{\citepads}[3][][]{\href{http://adsabs.harvard.edu/abs/#3}%
    {\def\hyper@linkstart##1##2{}%
     \let\hyper@linkend\@empty\citep[#1][#2]{#3}}}
  \newcommandtwoopt{\citetads}[3][][]{\href{http://adsabs.harvard.edu/abs/#3}%
    {\def\hyper@linkstart##1##2{}%
     \let\hyper@linkend\@empty\citet[#1][#2]{#3}}}
  \newcommandtwoopt{\citeyearads}[3][][]%
    {\href{http://adsabs.harvard.edu/abs/#3}
    {\def\hyper@linkstart##1##2{}%
     \let\hyper@linkend\@empty\citeyear[#1][#2]{#3}}}
\makeatother
\usepackage{todonotes}

\def\la{\mathrel{\hbox{\rlap{\hbox{\lower4pt\hbox{$\sim$}}}\hbox{$<$}}}}
\def\ga{\mathrel{\hbox{\rlap{\hbox{\lower4pt\hbox{$\sim$}}}\hbox{$>$}}}}

\def\deg      {{\ifmmode^\circ\else$^\circ$\fi} } 
\def\arcmin   {{\ifmmode {'}\else$'$\fi}} 
\def\arcsec   {{\ifmmode{''}\else$''$\fi}} 

\begin{document} 
   \title{Dynamical friction on an extended satellite revisited}
   \author{Pierfrancesco Di Cintio
          \inst{1,2,3}
          \and
         Carlo Nipoti
          \inst{4}
          }
   \institute{Consiglio Nazionale delle Ricerche, Istituto dei Sistemi Complessi, via Madonna del Piano 17, 50022 Sesto Fiorentino (FI), Italy\\
              \email{pierfrancesco.dicintio@cnr.it}
         \and
             INFN -- Sezione di Firenze, via G. Sansone 1, 50022 Sesto Fiorentino (FI), Italy
         \and
             INAF -- Osservatorio Astrofisico di Arcetri, Largo E. Fermi 5, 50125 Firenze, Italy   
         \and
               Dipartimento di Fisica e Astronomia “Augusto Righi” – Alma Mater Studiorum – Università di Bologna, via Gobetti 93/2, I-40129 Bologna\\
            \email{carlo.nipoti@unibo.it}
             }
 \date{Draft, \today}
  \abstract
   {The original formulation of Chandrasekhar dynamical friction (DF) was developed for a point-like particle moving in a infinite and homogeneous medium. However, despite  very useful order of magnitude estimates of the slowing-down time scales of a test mass in more realistic finite-sized models, high resolution $N-$body simulations often hint at a breakdown of Chandrasekhar's formalism (typically overestimating the effect of DF) for massive satellites sinking in the host galaxy.}
   {More advanced semi-analytical treatment of the DF expression managed to partially reduce said discrepancy, even though an accurate comparison with numerical simulations still remains difficult, specifically due to the interplay between the DF force and the tidal stress effects on an extended stellar systems sinking in the gravitational potential of the parent galaxy or halo.}
   {In this study we perform ad hoc $N-$body simulations of a satellite orbiting in a simple spherical galaxy model, for different values of the initial orbital eccentricity, satellite mass, scale radius and velocity anisotropy profile of the host model. We adopt an $N-$body code where the self-consistent interaction among the particles representing the galaxy can be substituted by the effect of the static smooth potential of the system's continuum limit. Moreover, the satellite itself is either modeled with $N_{\rm sat}$ particles or represented by a single particle with an effective size parametrized in the form of the force exchanged with the other simulation particles representing the galaxy.}
   {We find that, at fixed galaxy model and satellite mass, the more extended the satellite, the weaker is the effect of DF. This trend, which is found for both isotropic and radially anisotropic host systems, is particularly evident for orbits with initial orbital eccentricity exceeding  $\approx0.3$. }
   {}
   \maketitle
\section{Introduction}
Dynamical friction (hereafter DF) is the collective process that drives the orbital decay of a massive particle $m_t$ in the host stellar system. In its original inception (\citealt{1943ApJ....97..255C,1943ApJ....97..263C,1943ApJ....98...54C}), the effective drag is ascribed to (assumed) independent deflections of lighter field particles with typical mass $m$. Assuming an infinitely extended background system with fixed constant number and mass density $n$ and $\rho=mn$ with isotropic velocity distribution $f(v)$, the DF force term reads
\begin{equation}\label{eq:simpleDF}
\frac{{\rm d}\mathbf{v}}{{\rm d}t}=-\eta\mathbf{v}=-4\pi G^2\rho(m_t+m)\ln\Lambda\frac{\Xi(v)}{v^3}\mathbf{v}.
\end{equation}
The quantity $\ln\Lambda$, in analogy with the equivalent concept in plasma physics, is the so-called Coulomb logarithm of the maximum to minimum impact parameters $\Lambda =b_{\rm max}/b_{\rm min}$ in the scattering driven by $1/r^2$ forces. For the majority of the systems of astrophysical interest one has $\log\Lambda\approx 10$ and typically sets the factor $(m_t+m)$ equal to $m_t$ in the limit $m_t\gg m$. The fractional velocity volume function (e.g. see \citealt{binney,2021isd..book.....C})
\begin{equation}\label{eq:velfunct}
\Xi(v)=4\pi \int_0^{v}f(v^\prime)v^{\prime 2}{\rm d}v^\prime
\end{equation}
instead counts the portion of field particles $m$ moving slower than $m_t$ and thus contributing to DF for a given velocity distribution $f(v)$.\\
\indent When treating a finite system, typically, and rather arbitrarily (e.g. see the discussion in \citealt{1984MNRAS.209..729T}), the constant background density $\rho$ appearing in Equation (\ref{eq:simpleDF}) is substituted with the local density of the host system, while $f(v)$ is assumed to be a Maxwell-Boltzmann distribution. Despite its manifest simplicity, and the strong assumptions listed above, the classical DF formula has been proved successful over a broad range of astrophysical systems in estimating the infall time scales of massive stars in globular clusters (\citealt{1987degc.book.....S}) or dwarf satellites in the parent galaxy or dark matter halo. As expected, numerical simulations have evidenced some regimes where a direct application of Eq. (\ref{eq:simpleDF}) leads to either underestimate or overestimate the DF effect on a test particle. For instance, \cite{2006MNRAS.373.1451R} and \cite{2006MNRAS.368.1073G,2010ApJ...725.1707G} noticed that a massive test particle initially set on a circular orbit in a host model characterized by a centrally cored density profile after a ''super Chandrasekhar" phase\footnote{\cite{2016MNRAS.463..858P} interpreted the enhancement of DF as a consequence of factorizing the density and the velocity distribution and using a local Maxwellian approximation to evaluate $\Xi(v)$, rather than using the self consistent phase-space distribution $f$ of the host model.} (i.e. the orbit decays more rapidly that predicted by Eq. \ref{eq:simpleDF}), ''stalls" at a critical radius $r_s$ depending on the ratio $\alpha$ between the mass of the test particle and the host system, the resolution and, possibly, the orbital structure of the model. The hampering of the DF in cored systems\footnote{Interestingly, \cite{2003A&A...405...73B,2006A&A...453....9A,2007A&A...463..921A}, when accounting for the feedback on the background density found that the simple DF formula works well in models with broad density cores, while it fails in highly concentrated ones for which the feedback becomes more and more negligible.}, typically referred to as core-stalling, was also observed by \cite{2016MNRAS.455.3597Z} in numerical simulations of dwarf galaxies  sinking in flat galactic cores, and by \cite{2018ApJ...868..134K}, \cite{2019ApJ...877..133D} and \cite{2024A&A...690A.119B} when simulating the dynamics of globular clusters in galactic satellites. and by \cite{2024MNRAS.534..957G} in simulations of supermassive black holes (SMBHs) in nuclear star clusters.\\ 
\indent \cite{2021ApJ...912...43B,2022ApJ...926..215B,2025arXiv251111804D} explained the partial suppression of DF invoking the collisionless \cite{1972MNRAS.157....1L} torque, generalized relaxing the secular and adiabatic assumptions thus accounting for the contribution of the non-resonant orbits to the slowing down of the massive test particle. In this frame, once the test mass enters the core, the structure of the relative orbits of the near resonant field particles $m$ with respect to $m_t$ (e.g. see \citealt{2011MNRAS.416.1181I}) changes significantly. Since the phase space distribution in nearly harmonic cores is characterized by rather gentle gradients, the net torque from said orbits is statistically almost always positive, thereby countering the friction (an effect labeled as dynamical buoyancy). The core-stalling is therefore happening once the torque induced buoyancy balances DF\footnote{Actually, higher order resonance with non-corotating particles still induce a weak DF (\citealt{2022MNRAS.515..407K}), rendering a perfect core stalling impossible.}. \cite{2025A&A...700A.230D} on the other hand interpreted the stalling as a resolution-dependent effect of the discrete force fluctuations countering the DF.\\ 
\indent The original Chandrasekhar formulation of DF has been since extended to account for relativistic velocities (\citealt{1969ApJ...155..687L,1994MNRAS.270..205S,2017PhRvD..95f4014C,2023A&A...677A.140C}), non-spherical (\citealt{1977MNRAS.181..735B,1988ApJ...331...71S,1991ApJ...375..544S,1992MNRAS.254..466P,2002MNRAS.333..779P,2004MNRAS.349..747P,Vicari_2007,2022MNRAS.510.5945M}) and rotating hosts (\citealt{1986ApJ...306..374K,2018MNRAS.473.1719L}), multi-mass systems (\citealt{2010AIPC.1242..117C,2014ApJ...795..169A}) as well as alternative gravities (\citealt{2004MNRAS.351..285C,2006MNRAS.370.1829S,2008MNRAS.386.2194N,2024A&A...689A.150D,2026MNRAS.548ag017M}). Observations on numerical simulations revealed that stellar systems are in general far from being supported by isotropic velocity distributions. Surprisingly, relatively few studies have been devoted to investigate the effect of the specific form of the velocity distribution (beyond the isotropic Maxwellian approximation) or its degree of anisotropy. For example, we recall that \cite{2015ApJ...799...44M} derived Eq. (\ref{eq:velfunct}) in a \cite{1911MNRAS..71..460P} sphere integrating over its ergodic phase-space distribution $f(\mathcal{E})$ instead of its bona-fide factorization $\rho(r)f(v)$, while \cite{2016JCAP...05..021S} considered multi power-law velocity distributions in lieu of the standard Maxwellian approximation.\\
\indent The effect of the nonzero extent of the test body on the DF exerted by the host system has been mostly explored in the context where the test body is a rotating bar in a galactic disk (\citealt{1985MNRAS.213..451W,1998ApJ...493L...5D,1999MNRAS.306..662T,2006ApJ...637..567S,2008MNRAS.386.2194N,2023MNRAS.521.2227B,2024MNRAS.528.4115C,2026arXiv260526225G}), while relatively few works have systematically investigated the role of the test object's size when it is a satellite. In particular, \cite{1976MNRAS.174..467W} reworked Eq. (\ref{eq:simpleDF}) simply assuming that $m_t$ is a finite sized undeformable object. This substantially imposes an effective $b_{\rm min}$ in the definition of $\log\Lambda$ of the order of the scale size of the test mass itself (see e.g. \citealt{1998ApJ...502..150C}). \cite{1983A&A...117....9M} derived an analytical expression for an extended object with density $\rho_{\rm obj}(r)$ propagating through an infinite and uniform medium with density $\rho_0$ while accounting for both the response on the background density and the deformation of the test object. In this case, the total friction force is obtained by integral averaging the DF force acting on the the satellite's volume element $\mathbf{F}_z(R,z)$ over $\rho_{\rm obj}$ in cylindrical coordinates $R,z,\phi$ as
\begin{equation}\label{eq:mulder}
\frac{{\rm d}\mathbf{v}}{{\rm d}t}=2\pi\int_{-z_2}^{z_2}\int_0^{R_2}\mathbf{F}_z(R,z)\rho_{\rm obj}[(R^2+z^2)^{1/2}]R{\rm d}R{\rm d}z,
\end{equation}
In the formula above, the extremes of integration $z_2$ and $R_2$ are model dependent and they should be accommodated according to the specific form of $\rho_{\rm obj}$ (see Appendix II in \citealt{1983A&A...117....9M}).\\
\indent More recent studies on the dynamics of satellites in galactic potentials typically rely on point-particle estimations of the DF, often assuming isotropic velocity distributions, somewhat overlooking how much the intrinsic size of the satellite and the galactic or halo orbital structure could affect it. In this work we investigate this matter further, using $N-$body simulations of galaxy models with tunable degrees of velocity anisotropy as well as semi-analytical orbit propagation. The rest of the paper is structured as follows: In Sect. \ref{sec:models} we introduce the models used for the host galaxy and the satellites, the implementation of Eq. (\ref{eq:simpleDF}) and the numerical codes used in the simulations, then presented in Sect. \ref{sec:simu}. In Sect. \ref{sec:implication} we discuss the astrophysical implications of our findings and compare them with previous work. Finally, in Sect. \ref{sec:summary} we draw our conclusions.
\section{Models and methods}\label{sec:models}
\subsection{Galaxy and satellite models}
In the numerical simulations discussed in this work we model the host galaxy with a spherical \cite{1990ApJ...356..359H} model 
 with density distribution given by
\begin{equation}\label{eq:hernquistrho}
\rho_{\rm gal}(r)=\frac{1}{2\pi}\frac{M_{\rm gal}r_{\rm gal}}{r(r+r_{\rm gal})^{3}},
\end{equation}
and gravitational potential
\begin{eqnarray}\label{eq:hernquistphi}
 \Phi_{\rm gal}(r)=-\frac{GM_{\rm gal}}{r+r_{\rm gal}}.
 \end{eqnarray} 
For the satellite we adopt the \cite{1911MNRAS..71..460P} model with density
\begin{equation}\label{eq:plummerrho}
    \rho_{\rm sat}(r)=\frac{3 M_{\rm sat}}{4\pi r_{\rm sat}^3}\left( 1 + \frac{r^2}{r_{\rm sat}^2} \right)^{-5/2}
\end{equation}
and potential
\begin{equation}\label{eq:plummerphi}
   \Phi_{\rm sat}(r)=-\frac{GM_{\rm sat}}{\sqrt{r^2+r_{\rm sat}^2}}.
\end{equation}
The initial conditions for the $N-$body simulations are implemented as follows. First $N_s$ equal particles representing the host galaxy are sampled with the usual Monte-Carlo scheme after inverting the radial mass profile
\begin{equation}\label{eq:mr}
M_{\rm gal}(r)=M_{\rm gal}\left(\frac{r}{r+r_{\rm gal}}\right)^{2}.
\end{equation}
Then, to account for radially biased velocity anisotropy profile, the particle velocities are drawn using the standard rejection method (for the implementation details see e.g. \citealt{2015JPlPh..81e4904D}) from a Osipkov-Merritt (hereafter, OM; \citealt{1979SvAL....5...42O,1985AJ.....90.1027M}) distribution function given by 
\begin{equation}\label{eq:OM}
f(Q)=\frac{1}{\sqrt{8}\pi^2}\int_Q^{0}\frac{{\rm d}^2\rho_a}{{\rm d}\Phi^2}\frac{{\rm d}\Phi}{\sqrt{\Phi-Q}}.
\end{equation}
In Equation (\ref{eq:OM}) 
\begin{equation}\label{eq:Q}
Q=E+{J^2}/{2r_a^2}, 
\end{equation}
 $E$ and $J$ are the particle's energy and angular momentum per unit mass, and 
 \begin{equation}\label{eq:augmented}
 \rho_a(r)\equiv\left(1+{r^2}/{r_a^2}\right)\rho(r)
 \end{equation}
is the augmented density. The parameter $r_a$, the so-called anisotropy radius, controls the extent of the radial velocity anisotropy: for $r\ll r_a$ the model is nearly isotropic, while for $r\gg r_a$ the velocity distribution is dominated by low angular momentum, nearly radial orbits. The total amount of anisotropy is quantified with the usual Fridman-Polyachenko-Shukhman (\citealt{1984sv...bookR....F}; see also \citealt{binney})
\begin{equation}\label{eq:xi}
 \xi\equiv\frac{2K_r}{K_t},
\end{equation}
where the radial and tangential kinetic energies are given respectively by
\begin{equation}\label{eq:krkt}
K_r=2\pi\int\rho(r)\sigma^2_r(r)r^2{\rm d}r,\quad K_t=2\pi\int\rho(r)\sigma^2_t(r)r^2{\rm d}r.
\end{equation}
For the satellite, for which the radial mass profile is given by
\begin{equation}\label{eq:mrsat}
M_{\rm sat}(r)=M_{\rm sat}\left(\frac{r^2}{r^2+r_{\rm sat}^2}\right)^{3/2},
\end{equation}
we always use the isotropic phase-space distribution function 
\begin{equation}\label{eq:fe}
f(E)=\frac{\sqrt{2}}{378\pi^2Gr_{\rm sat}^2\sigma_0}\left(\frac{-E}{\sigma_0^2}\right)^{7/2},
\end{equation}
where $\sigma_0=\sqrt{GM_{\rm sat}/6r_{\rm sat}}$ is the one dimensional velocity dispersion at $r=0$.\\
\indent When dealing with extended systems orbiting in external potentials, it is worth establishing how their effective size compares with their tidal radius $r_T$, which, following
 \cite{1962AJ.....67..471K},  for a satellite of mass  $M_{\rm sat}$ orbiting a host with mass $M_{\rm gal}$, can be defined as
\begin{equation}\label{eq:rtide}
r_T=d\left(\frac{M_{\rm sat}}{\beta M_{\rm gal}}\right)^{1/3},
\end{equation}
where the dimensionless factor $\beta$ equals 3 (regardless the orbital eccentricity of $M_{\rm sat}$) in the Keplerian potential of a point-like particle body of mass $M_{\rm gal}$, while it becomes 2 for an extended host wherein $M_{\rm sat}$ moves at constant speed $v_c$ on a circular orbit (\citealt{1990ApJ...351..121C}).
\begin{figure}
	\centering
	\includegraphics[width=0.9\columnwidth]{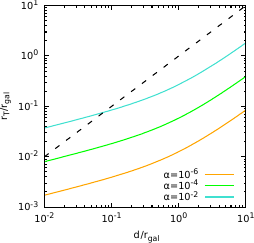}
	\caption{Tidal radius as a function of the galactocentric distance $d$ for a satellite of mass $M_{\rm sat}$ orbiting in a Hernquist galaxy model of mass $M_{\rm gal}$ and radius $r_{\rm gal}$. The different coloured solid curves refer to different choices of the mass ratio $\alpha=M_{\rm sat}/M_{gal}$. To guide the eye, the dashed curve divides the plot in a region where $r_{T}>d$ and $r_{T}<d$.}
	\label{fig_rtide}
\end{figure}
In this work we consider a modified version of Eq. (\ref{eq:rtide}) where we substitute $M_{\rm gal}$ with $M_{\rm gal}(r)$ evaluated instantaneously at $d$ according Eq. (\ref{eq:mr}) and we set $\beta=2$. In Figure \ref{fig_rtide} we show the associated dependence of $r_T$ on the galactocentric distance for three values of the ratio $\alpha=M_{\rm sat}/M_{\rm gal}=10^{-6}$, $10^{-4}$ and $10^{-2}$. 
\subsection{$N-$body method and simulation setup}
\label{sec:nbody_method}
In this work we performed three different kinds of $N-$body experiments, which we define here below. 
\begin{enumerate}
\item "Full $N-$body simulations", in which the host galaxy and the satellite are represented, respectively, by $N_{\rm gal}$ and $N_{\rm sat}$ equal mass simulation particles.
\item "Semi-reduced $N-$body simulations", in which the galaxy is still modelled with $N_{\rm gal}$ interacting particles, while the satellite is represented by a single extended particle of mass $M_{\rm sat}$ that interacts directly with the $N_{\rm gal}$ simulation particles representing the galaxy via a Plummer potential (\ref{eq:plummerphi}) with scale length corresponding to the specific choice of $r_{\rm sat}$. In this approach we neglect all the effects related to the deformation of the satellite and/or tidal stripping.
\item "Reduced $N-$body simulations", in which we further simplify the simulation set-up by propagating the $N_{\rm gal}$ non-interacting particles representing the host in the parent smooth potential given by Eq. (\ref{eq:hernquistphi}), in a similar fashion to what done in previous studies (see e.g. \citealt{1983ApJ...274...53W,1987MNRAS.224..349B,1989MNRAS.238..407H,2020IAUS..351..532V,2025A&A...700A.230D} and references therein), and letting then exert their force on the extended particle of mass $M_{\rm sat}$ modelling the satellite.
\end{enumerate}
 \indent In the full $N-$body and semi-reduced simulations all forces are evaluated with the fast \cite{2000ApJ...536L..39D} tree scheme implemented in the {\sc fvfps} code (see \citealt{2003MSAIS...1...18L,2006MNRAS.370..681N}), and a fixed softening length $\epsilon_s=3\times10^{-3}r_{\rm gal}$ regularizes interactions among nearby particles. In in the reduced and semi-reduced simulations we evaluate the force due to the model particles $m$ acting on the satellite at position $\mathbf{d}$ in the galactocentric frame by direct sum as
 \begin{equation}\label{eq:forcesat}
\mathbf{a}_{\rm sat}=-Gm\sum_{i=1}^{N_{\rm gal}}\frac{(\mathbf{r}_i-\mathbf{d})}{(r_{\rm sat}^2+||\mathbf{r}_i-\mathbf{d}||^2)^{3/2}}.
 \end{equation}
 In the full $N-$body simulations discussed here we always use the same resolution for the host and its satellite, that is, $M_{\rm gal}/N_{\rm gal}=M_{\rm sat}/N_{\rm sat}$ and all simulation particles have equal masses $m$. In practice, when generating the $N-$body realization of the satellite Eq.(\ref{eq:mrsat}) is multiplied by  $[(r_*^2/(r_*^2+r^2)]^{-3/2}$, where $r_*$ is a truncation radius (such that $\rho_{\rm sat}=0$ for $r\geq r_*$).  As a rule (except where explicitly indicated), we truncate the radial density distribution of the satellite at $r_*=3r_{\rm sat}$.\\
\indent In all sets of simulations introduced here $M_{\rm sat}$ ranges from $10^{-3}$ to $10^{-2}$ in units of $M_{\rm gal}$. Except where otherwise indicated, we fix $N_{\rm gal}=10^6$ and chose $N_{\rm sat}$ accordingly. We initialized the satellite on orbits with eccentricities $e$ from $\approx0$ to roughly 0.9 corresponding to nearly circular to relatively eccentric orbits, respectively. As usual the eccentricity is defined as
\begin{equation}\label{eq:ecc}
e = \frac{r_{\text{apo}} - r_{\text{peri}}}{r_{\text{apo}} + r_{\text{peri}}},
\end{equation}
where $r_{\text{apo}}$ and $r_{\text{peri}}$ are the apocentre and pericentre of the unperturbed orbit in the smooth potential $\Phi$, respectively.\\
\indent All particles orbits are propagated with a second order leapfrog scheme with fixed time step $\Delta t=t_{c}/150$, where the reference time scale is set to
\begin{equation}\label{eq:tdyn}
t_c=\sqrt{r_{\rm gal}^3/GM_{\rm gal}},
\end{equation}
that is of the order crossing time at the half-mass radius.
\begin{figure}
	\centering
	\includegraphics[width=0.9\columnwidth]{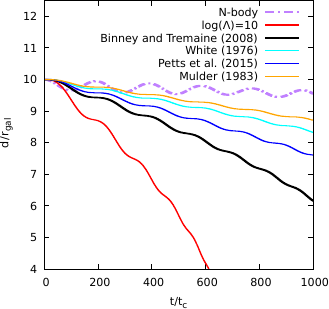}
	\caption{Radial decay of satellite with $M_{\rm sat}=10^{-3}M_{\rm gal}$ and $r_{\rm sat}=10^{-1}r_{\rm gal}$ orbiting in a Hernquist model evaluated applying the Chandrasekhar formula (Eq. \ref{eq:simpleDF}) with different choices of $\log{\Lambda}$ indicated by the red ($\log\Lambda =10$), cyan (\citealt{1976MNRAS.174..467W}), black (\citealt{binney}),  blue (\citealt{2015MNRAS.454.3778P}) and yellow (\citealt{1983A&A...117....9M}) solid lines, respectively. The dot-dashed purple indicates the decay of the galactocentric distance $d$ in a reduced $N-$body simulation where the satellite is represented by a single extended particle.}
	\label{fig_lambda}
\end{figure}
\subsection{Semi-analytical models}
For comparison with the simulations described in \ref{sec:nbody_method}, we have run additional calculations, which we refer to as "semi-analytical models", in which the same initial conditions for the satellite orbit considered in the $N$-body simulations (Section \ref{sec:nbody_method}) were propagated in the smooth galactic potential given by Eq.\ (\ref{eq:hernquistphi}) under the effect of the dynamical friction as predicted by the Chandrasekhar formula. In practice, the equations of motion for the particle representing the satellite in the galactic potential are augmented by the DF term of Eq.\ (\ref{eq:simpleDF}) as
\begin{equation}\label{eq:eom}
\frac{{\rm d}^2\mathbf{r}}{{\rm d}t^2}=-\nabla\Phi_{\rm gal}-\eta(r,v)\mathbf{v},
\end{equation}
where $\eta(r,v)$ is the position and velocity dependent DF coefficient.\\
\indent In line with previous work, we accommodate the definition of the DF force using the local values of $\rho$ (e.g. see the discussion in \citealt{doi:10.1137/050646032}) and we adopt the typical local Maxwellian approximation on $f(v)$ (i.e. at each radius the velocity distribution is a isotropic Maxwellian with standard deviation equal to the local velocity dispersion $\sigma$ ) such that, see Eq. (\ref{eq:velfunct}),
\begin{equation}\label{eq:localmax}
\Xi(v)={\rm Erf}\left(\frac{v}{\sqrt{2}\sigma}\right)-\sqrt{\frac{2}{\pi}}\frac{v}{\sigma}\exp\left(-\frac{v^2}{2\sigma^2}\right).
\end{equation}
For our choice of host density profile (\ref{eq:hernquistrho}), $\sigma$ is known analytically (\citealt{1990ApJ...356..359H}). Since the equation of motion (\ref{eq:eom}) for $M_{\rm sat}$ contains a velocity-dependent dissipative term, we do not adopt the same symplectic scheme used for the conservative $N-$body simulation. We use instead a modified mid-point leapfrog, typically adopted for the integration of orbits with velocity dependent force terms, see for example \citealt{mik06} for the details.\\
\indent The definition of the Coulomb logarithm's argument is a rather delicate point, even in the the simple formulation (\ref{eq:simpleDF}). Whereas the minimum impact parameter can be set to its critical value for a $\pi/2$ deflection at typical speed $v_t$ as $b_{\rm min}=Gm_t/v_t^2$ (\citealt{binney}), the value of $b_{\rm max}$ remains somewhat arbitrary (e.g. see \citealt{1983Ap&SS..97..435K}) and it is often set to the galaxy scale radius $r_{\rm gal}$. For the case of an extended object, following the original \cite{1976MNRAS.174..467W} assumption,\footnote{Other more recent numerical studies (\citealt{2012ApJ...753....9V, 2024A&A...685A..38C}) adopted a similar choice of $\log\Lambda$, tailored on $N-$body simulations, see also \cite{2003ApJ...582..196H}.} one therefore set $\log\Lambda=\log (r_{\rm gal}/r_{\rm sat})$. 
\cite{2015MNRAS.454.3778P,2016MNRAS.463..858P} (see also \citealt{2005A&A...431..861J,2019MNRAS.484.5409P}) proposed the more elaborate local density gradient-dependent recipe  
\begin{equation}\label{eq:petts+}
\log\Lambda=\log\frac{{\rm min}(\rho_{\rm gal}/||\nabla\rho_{\rm gal}||;d)}{{\rm max}(r_h;GM_{\rm sat}/v^2)},
\end{equation}
where $d$ is the galactocentric distance of the satellite and $r_h$ its half mass radius, which for our choice of the satellite's density profile\footnote{We recall that orbit integration software such as {\sc galpy} (\citealt{2015ApJS..216...29B}) can substitute ${\rm min}(\rho_{\rm gal}/||\nabla\rho_{\rm gal}||;d)$ in Eq. (\ref{eq:petts+}) with $d/\gamma$ where $\gamma$ is the (local) logarithmic density slope. Clearly, in cored systems where $\gamma\rightarrow 0$ in the central regions this simplification is not applicable.} (\ref{eq:plummerrho}) is $r_h=r_{\rm sat}/(\sqrt{2}-1)$.\\
\indent We tested for various orbital parameters in the potential (\ref{eq:hernquistphi}) the implementations of the Coulomb logarithm in Eq. (\ref{eq:simpleDF}) discussed by \cite{1976MNRAS.174..467W} and \cite{2015MNRAS.454.3778P} as well as the formulation given by \cite{binney} and a constant $\log\Lambda=10$.
\begin{table*}
\caption{Satellite orbit's initial conditions and model properties}
\centering
\begin{tabular}{llllllll}
\hline
Simulation type  & $M_{\rm sat}/M_{\rm gal}$  & $r_{\rm sat}/r_{\rm gal}$  & $N_{\rm gal}$  & $N_{\rm sat}$       & $\xi_{\rm gal}$         & $d_{0}/r_{\rm gal}$  & $e_0$    \\ \hline
(1)    & (2)   & (3)    & (4)   &  (5)     & (6)  & (7)  & (8)       \\ \hline
Full $N-$body    & $10^{-3}$  & $10^{-2}$   & $10^6$  &  $10^3$    & 1 & 10 & 0.9      \\ 
Full $N-$body    & $10^{-2}$  & $10^{-1}$   & $10^6$  &  $10^4$    & 1 &  6 & 0.24      \\ 
Full $N-$body    & $10^{-3}$  & $10^{-2}$   & $10^6$  &  $10^3$    & 1 &  2 & 0.13      \\ 
Full $N-$body    & $3\times10^{-3}$  & $8.5\times10^{-2}$   & $10^6$  &  $3\times10^3$    & 1 &  5 & 0.04      \\
Semi-analytical    & $10^{-3}$  & $10^{-2}$   & $/$  &  $1$    & 1 & 10 & 0.9      \\ 
Semi-analytical    & $10^{-2}$  & $10^{-1}$   & $/$  &  $1$    & 1 &  6 & 0.24      \\ 
Semi-analytical    & $10^{-3}$  & $10^{-2}$   & $/$  &  $1$    & 1 &  2 & 0.13      \\ 
Semi-analytical    & $3\times10^{-3}$  & $8.5\times10^{-2}$   & $/$  &  $1$    & 1 &  5 & 0.04      \\
Semi-reduced $N-$body   & $10^{-3}$, $10^{-2}$  & $10^{-2}$, $5\times 10^{-2}$   & $10^6$  &  1    & 1, 1.7 & 10 & 0.9      \\ 
Semi-reduced $N-$body    & $10^{-2}$  & $10^{-2}$, $5\times 10^{-2}$, $10^{-1}$   & $10^6$  &  1    & 1, 1.7 &  6 & 0.24      \\ 
Semi-reduced $N-$body    & $10^{-3}$, $10^{-2}$  & $10^{-2}$ , $5\times 10^{-2}$  & $10^6$  &  1    & 1, 1.7 &  2 & 0.13      \\ 
Semi-reduced $N-$body    & $3\times10^{-3}$, $10^{-2}$  & $10^{-2}$, $5\times 10^{-2}$, $8.5\times10^{-2}$   & $10^6$  &  1    & 1, 1.7 &  5 & 0.04     \\ 
Reduced $N-$body    & $10^{-3}$  &  $10^{-3}$, $10^{-2}$, $10^{-1}$   & $10^6$  &  1    & 1 & 10 & 0.02      \\
Reduced $N-$body    & $10^{-2}$  &  $10^{-3}$, $10^{-2}$, $10^{-1}$   & $10^6$  &  1    & 1 & 10 & 0.02      \\
Semi-analytical    & $10^{-3}$, $10^{-2}$  & /   & /  &  1    & 1 & 10 & 0.02      \\
Reduced $N-$body    & $10^{-3}$  &  $10^{-3}$, $10^{-2}$, $10^{-1}$   & $10^6$  &  1    & 1 & 6 & 0.24      \\
Reduced $N-$body    & $10^{-2}$  &  $10^{-3}$, $10^{-2}$, $10^{-1}$   & $10^6$  &  1    & 1 & 6 & 0.24      \\
Semi-analytical    & $10^{-3}$, $10^{-2}$  & /   & / &  1    & 1 & 6 & 0.24      \\
Reduced $N-$body    & $10^{-3}$  &  $10^{-3}$, $10^{-2}$, $10^{-1}$   & $10^6$  &  1    & 1 & 10 & 0.9      \\
Reduced $N-$body    & $10^{-2}$  &  $10^{-3}$, $10^{-2}$, $10^{-1}$   & $10^6$  &  1    & 1 & 10 & 0.9      \\
Semi-analytical    & $10^{-3}$, $10^{-2}$  & /   & /  &  1    & 1 & 10 & 0.9      \\
\hline
\hline
\end{tabular}
\tablefoot{Simulation type (1), mass ratio (2), ratio of scale radii (3), number of particles used for the host galaxy (4) and the satellite (5), anisotropy index (6), initial galactocentric distance (7) and initial orbital eccentricity (8).}
\label{tab:init}
\end{table*}
As an example, in Figure \ref{fig_lambda} the different lines show the estimated decay of the galactocentric distance of a satellite with $M_{\rm sat}=10^{-3}M_{\rm gal}$ and $r_{\rm sat}=10^{-1}r_{\rm gal}$ initially placed on a nearly circular orbit of apocentric radius $r_0=10r_{\rm gal}$ in a Hernquist galaxy model. For comparison, the evolution of $d$ in a semi-reduced $N-$body simulation is marked by the purple heavy dotted-dashed line. Remarkably, for this specific set-up, all orbits estimated using the classical DF expression decay much faster than the parent case in the semi-reduced $N-$body simulation, regardless of the choice of $\Lambda$ at hand. 
\begin{figure*}
	\centering
	\includegraphics[width=\textwidth]{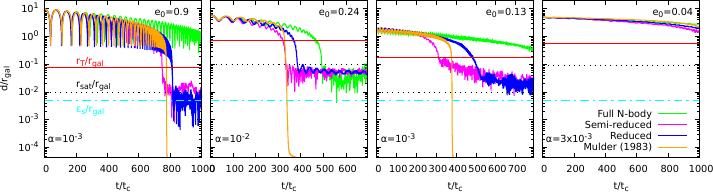}
	\caption{Radial decay of a satellite orbiting a Hernquist model for, respectively, from left to right, mass ratio $\alpha=M_{\rm sat}/M_{\rm gal}=10^{-3}$, $10^{-2}$, $10^{-3}$, $3\times 10^{-3}$, initial orbital eccentricity $e_0=0.9$, 0.24, 0.13 and 0.04, and satellite-to-host scale radius ratio $r_{\rm sat}/r_{\rm gal}=10^{-2}$, $10^{-1}$, $10^{-2}$ and $8.5\times 10^{-2}$. In all panels the green lines correspond to the full $N-$body simulations, the magenta lines to the semi-reduced $N-$body simulations, the blue lines to the reduced $N-$body simulations and the orange curves to the semianalytical model (\ref{eq:mulder}). The red solid, black dotted and cyan dashed-dotted lines mark, respectively, the estimated tidal radius at the first pericenter, the initial values of $r_{\rm sat}/r_{\rm gal}$, and the fixed softening length $\epsilon_s=4.5\times10^{-3}r_{\rm gal}$ used in full $N-$body and semi-reduced simulations.}
	\label{fig_nbody}
\end{figure*}
\begin{figure*}
	\centering
	\includegraphics[width=0.9\textwidth]{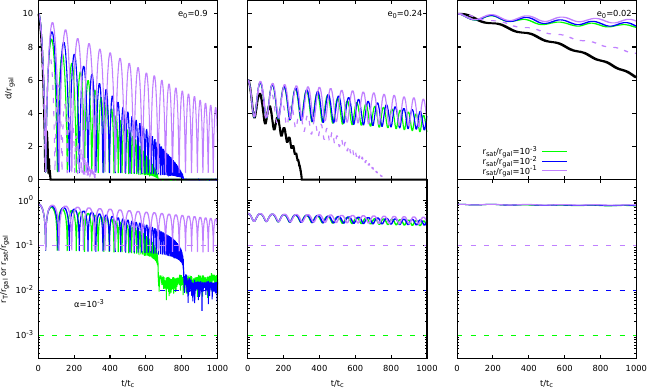}
	\caption{Top panels: Radial decay of a satellite with Plummer density profile and scale radius $r_{\rm sat}/r_{\rm gal}=10^{-3}$ (green lines), $10^{-2}$ (blue lines) and $10^{-1}$ (purple lines) with masses $M_{\rm sat}=10^{-3}M_{\rm gal}$, placed on an orbit with, from left to right, initial eccentricity $e_0=0.9$, 0.24 and 0.02 in a spherical Hernquist galaxy modeled with $N_{\rm gal}=10^6$ particles. The coloured solid lines refer to reduced $N-$body simulations while the dotted black and dashed purple lines mark the cases where the satellite was propagated in a smooth potential plus dynamical friction according to Eq. (\ref{eq:simpleDF}) with $\log\Lambda$ defined for a point-like mass \citep{binney} and for an extended satellite with $r_{\rm sat}=0.1r_{\rm gal}$ \citep{2015MNRAS.454.3778P}, respectively. Bottom panels: Evolution of the tidal radius $r_T$ (solid curves) and values $r_{\rm sat}$ (dashed curves) for the same simulations as in the corresponding upper panels (with the same color coding).}
	\label{fig_decaycirc}
\end{figure*}
 The simulations parameters for all the numerical experiments discussed in the present work are summarized in Tab. \ref{tab:init}.
\section{Numerical simulations}\label{sec:simu}
\begin{figure*}
	\centering
	\includegraphics[width=0.9\textwidth]{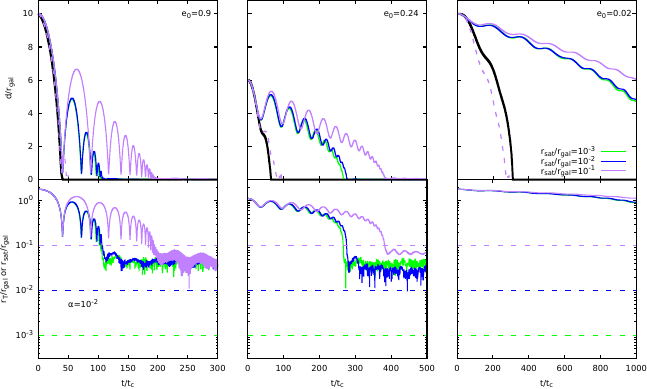}
	\caption{Same as in Fig. \ref{fig_decaycirc} for $M_{\rm sat}=10^{-2}M_{\rm gal}$.}
	\label{fig_decayecc_a}
\end{figure*}
\begin{figure*}
	\centering
	\includegraphics[width=\textwidth]{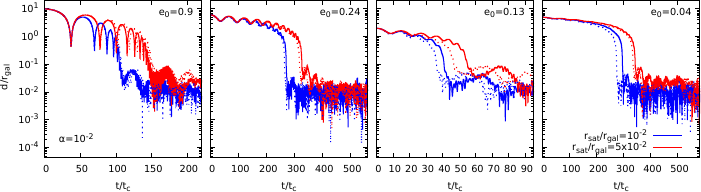}
	\caption{Radial decay of a satellite orbiting a anisotropic Hernquist model with $\xi=1.7$ in reduced simulations for $\alpha=M_{\rm sat}/M_{\rm gal}=10^{-2}$, and $r_{\rm sat}/r_{\rm gal}=10^{-2}$ (blue lines) and $5\times 10^{-2}$  (red lines), with initial orbital eccentricity, from left to right $e_0=0.9$, 0.24, 0.13 and 0.04. The thin dashed lines correspond to case of a satellite with the same parameters in the isotropic Hernquist host.}
	\label{fig_aniso}
\end{figure*}
\subsection{Different $N-$body set-ups}
We first compared the full $N-$body simulations, where both the galaxy and its satellite are modeled with particles with the semi-reduced and reduced experiments, where the satellite is represented instead by a single massive particle with an effective scale radius (larger than the softening length $\epsilon_s$ used in the simulation). In Figure \ref{fig_nbody} we present the evolution of the satellite galactocentric distance $d$ in four test cases where the satellite moves in an isotropic Hernquist galaxy model with (from left to right) decreasing initial orbital eccentricity $e_0=e(t=0)$ from 0.9 to 0.04. All runs were extended up to $t=1000t_c$ (corresponding in all cases to at least 20 pericentric passages) and start with the satellite at its orbital turning point at $r_{\rm apo}$. In all cases the combination of $e_0$ and $r_{\rm apo}$ is such that $r_{\rm T}$ at the first pericentric passage is larger than $r_{\rm sat}$. In the two middle panels the evolution of $d$ in the full $N-$body experiments (green curves) is shown only up to the time $t\approx 800t_c$, after which the satellite is almost completely disrupted by tidal forces. In the full $N-$body runs the galactocentric distance is evaluated as a function of time by estimating the position of the density peak of the $N_{\rm sat}$ particles representing the satellite.\\

\indent We observe that, at least up to $t\approx 200t_c$ (corresponding, respectively, to 3, 4, 15 and 5 pericentric passages in the panels from left to right of Figure \ref{fig_nbody}), independently of the mass ratio $\alpha$ and the value of $r_{\rm sat}$ all three simulation schemes yield the same evolution of the galactocentric distance. At later times, the response of the satellite's density to the tidal perturbations becomes relevant. In particular, we find that the orbital decay is significantly faster in the semi-reduced and reduced simulations (i.e. the DF is more effective) with respect to the the full $N-$body simulations, independent of the specific value of $M_{\rm sat}$ and $r_{\rm sat}$ at $e_0\gtrsim 0.1$. The satellites appears to stall near the same radius in simulations of different type. We run additional simulations in the reduced and semi-reduced configurations with different resolutions ranging from $N\approx 10^4$ to $N=3\times 10^6$. We find that, whereas the final stalling radius depends on $N$ (see the discussion in \citealt{2025A&A...700A.230D}), the radial decay down to at least $d\approx d_{0}/4$ is independent of the resolution for $N\gtrsim 5\times 10^4$.\\

\indent The systematically more effective friction acting on the satellite in both the semi-reduced and reduced cases, when compared with the fully self-consistent experiments, can be ascribed to the fact that $M_{\rm sat}$ is conserved throughout the whole simulation, while in a full $N-$body simulation the satellite can lose mass (e.g. see \citealt{2025arXiv250803286D,2025A&A...701A.129G}) thus effectively reducing the DF deceleration (see also \citealt{2018MNRAS.481.4863S}). Some authors, speculate that the so-called self-friction (hereafter SF), that is the drag exerted by long and massive tidal tails \citep{2020MNRAS.495.4496M} might induce an additional component to the drag felt by the test object. However, its magnitude is always rather small. We recall that SF is found to reach at most 10\% in magnitude of the DF standard term, typically this happens for satellites on highly elongated orbits or extremely bound ones. Among the simulations presented in this work, the satellite initially on the most eccentric $e_0=0.9$ orbit shows an almost conserved perigalactic distance in the full $N-$body run. We interpret this as due to the mass loss along a highly elongated orbit that makes the satellite considerably less massive. We note that \cite{2020MNRAS.495.4496M} finds that, for similar orbital eccentricity, the SF gives a slightly larger contribution the orbital decay. However, one must bear in mind that  different density profiles are used for the numerical experiments (in \citealt{2020MNRAS.495.4496M} both the host and the satellite are cuspy \citet{1996ApJ...462..563N} halos,  while here we use a Hernquist and a flat-cored Plummer profile, 
respectively) and the use of a live- or frozen-host model. In addition, the effective size of the satellite $r_{\rm sat}$ remains well below $r_T$, thus implying a limited extent of the tidal arms. We observe that, despite having the same stalling radius, in the semi-reduced $N-$body simulations the DF is always more efficient (i.e.\ shorter DF time) than in the reduced simulations, independently of the specific satellite model or orbital parameters. Additional experiments with different (lower) resolutions display the same behaviour. We conjecture that, the self-consistent wake (i.e.\ with its own self gravity and thus able to accrete extra mass) forming behind the fixed mass extended particle representing the satellite (e.g. see \citealt{2016MNRAS.457.2164O}) gives a bigger contribution to DF than in reduced simulations.\\
\indent For comparison, in Fig. \ref{fig_nbody} we also plot the modified DF expression of \cite{1983A&A...117....9M} (yellow curves), evaluated using a bona fide satellite density $\rho_{\rm obj}=3M_{\rm sat}/(8\pi r_h^3)$ in the assumption that roughly half of the satellite mass is always enclosed within its initial half mass radius $r_h$. This rather arbitrary choice was made since in the models considered here the estimated time dependent $r_T$ almost never reaches $r_{h}$. By doing so, the double integral in Eq.(\ref{eq:mulder}), carried out neglecting the $R$ and $z$ dependence inside $\log\Lambda$ (and therefore also taking out of the integral the factor $F_z(R,z)$), yields a corrective term to Eq. (\ref{eq:simpleDF}) in the form 
\begin{equation}\label{eq:alfa}
\mathcal{C}=\frac{3z_*R_*^2}{8r_{h}^3}, 
\end{equation}
\begin{figure}
	\centering
	\includegraphics[width=0.9\columnwidth]{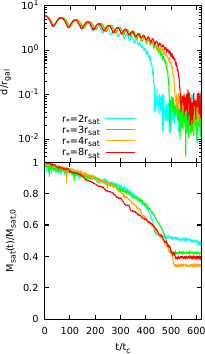}
	\caption{Evolution of the galactocentric distance for a $M_{\rm sat}=10^{-2}M_{\rm gal}$ satellite initially placed on a $e_0=0.24$ orbit (top panel), and corresponding time-dependent retained mass fraction (bottom panel) in full $N-$body simulations. Different curves refer to different values of the cut-off radius $r_*$ of the satellite's density. In all cases $r_{\rm sat}=0.1r_{\rm gal}$. The green curve represents the same simulation as the green curve in the second panel from the left in Fig.\ \ref{fig_nbody}.}
	\label{fig_cutoff}
\end{figure}
where we set $R_*={\rm min}(r_T;r_h)$ and $z_*={\rm min}(r_h;v^2/||\nabla\phi(d)||)$, being, respectively, the geometrical cross section of the satellite and the effective distance crossed between two independent scatters. The argument of the Coulomb logarithm $\Lambda$ has been assumed, following to \cite{2015ApJS..216...29B}, with $r_T$ in lieu of the half mass radius $r_{h}$. The evolution of $d$ estimated with the DF expression corrected by the factor $\alpha$ sits between the estimates of the reduced and semi-reduced simulations, which, as we pointed out above, do overestimate significantly the orbital decay. For the case of satellite on a low eccentricity orbit, Eq.\ (\ref{eq:mulder}) is closer to the satellite trajectory in the full $N-$body simulation.\\
\indent To summarize, it appears that, before substantial deformation or mass loss, the simpler and less time consuming reduced simulations can be used  as a probe of DF even for models of finite extent. In the following, with the exception of Sect. \ref{sec:rt}, we will restrict to this type of experiments.  
\begin{figure*}
	\centering
	\includegraphics[width=0.7\textwidth]{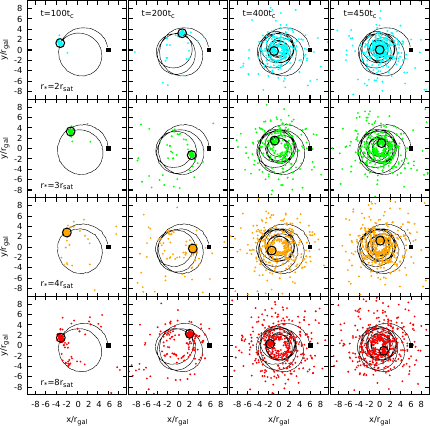}
	\caption{Particle distribution in the $x-y$ orbital plane at $t/t_c=100$, 200 and 400 for the models with truncation radii (from top to bottom) $r_*=2r_{\rm sat}$, $3r_{\rm sat}$ , $4r_{\rm sat}$ and $8r_{\rm sat}$ of Fig. \ref{fig_cutoff}. The thin black lines mark the trajectory of the satellite's density peak used to evaluate the galactocentric distance $d$, while the squares and circles mark the satellite's initial condition and its position at the indicated time, respectively.}
\label{fig_tides}
\end{figure*}
\begin{figure}
	\centering
	\includegraphics[width=0.9\columnwidth]{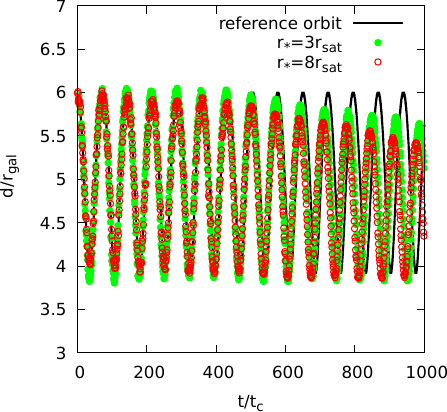}
	\caption{Orbital decay of $M_{\rm sat}=10^{-2}M_{\rm gal}$ satellite  with $r_{\rm sat}=0.1r_{\rm gal}$ initially placed on a $e_0=0.24$ orbit in a fixed Hernquist potential. The satellite is modelled as an $N$-body system with $r_*=3r_{\rm sat}$  (green circles) and $8r_{\rm sat}$ (red circles). DF is not included, so the orbital decay of the satellite is due only to SF with its tidal tails. The solid black line indicates the reference trajectory of a point mass in the smooth host potential.}
	\label{fig_self}
\end{figure}
\subsection{Dependence on the satellite scale radius $r_{\rm sat}$}
In order to extrapolate the dependence of the DF force on the satellite's scale radius $r_{\rm sat}$, we ran a set of reduced $N-$body simulations up to roughly 25 pericentre passages at fixed host properties and satellite mass, while changing only $r_{\rm sat}$ in the range ($10^{-3}r_{\rm gal}-10^{-1}r_{\rm gal}$). The top panels of Figures \ref{fig_decaycirc} and \ref{fig_decayecc_a} show the evolution of the galactocentric distance $d$ for a $10^{-3}M_{\rm gal}$ (left) and $10^{-2}M_{\rm gal}$ (right) satellite, initially on a nearly circular, moderately eccentric and highly eccentric orbit. In all cases, using the simple semi-analytical estimate of the DF for a point particle leads to overestimate the orbital decay (see the heavy black curves in the plots). The specific value of $r_{ \rm sat}$ has little effect on the strength of the drag when the orbital eccentricity is as low as 0.02, but for larger values of $e_0$, we observe as a general trend that the DF becomes less effective for more extended satellites for a given $M_{\rm sat}$. We performed additional semi-reduced $N-$body runs (not shown here) where the satellite is placed initially at $d_{0}\approx r_{\rm gal}$ on a nearly circular orbit. In general, for smaller values of $r_{\rm sat}$ we find an initially more efficient DF in the $N-$body simulation with respect to the semi-analytical estimate, always followed by a sub-Chandrasekhar regime (i.e. DF less efficient than its parent semi-analytical estimate), coherently with the findings of \cite{2006MNRAS.368.1073G} and \cite{2006MNRAS.373.1451R}.\\
In the lower panel of Figs. \ref{fig_decaycirc} and \ref{fig_decayecc_a}, we show the evolution of the tidal radius $r_T$ as a function of time (solid lines) and, for comparison, the values of the satellite scale radius (dashed lines). Only the models with the largest extent at fixed mass (i.e. $r_{\rm sat}=10^{-1}r_{\rm gal}$) on eccentric orbits decay to radii where $r_T<r_{\rm sat}$ thus rendering questionable the mass conserving extend particle approximation. However, we stress the fact that, for our purposes, this happens long after (i.e. $\approx10$ pericentric passages) the more concentrated models of the same mass have sunken in the central regions of the host. 
\subsection{Effect of the host velocity distribution}
The dynamics of a satellite sinking in the parent galaxy is likely to be affected by the host's orbital structure. Most numerical and analytical works on DF, when dealing with a spherical host, have assumed a isotropic velocity distribution. However, there is evidence, both from observations (e.g. \citealt{2007MNRAS.379..418C,2019MNRAS.487.3693J} and references therein) and $N-$body simulations (e.g.  \citealt{2006MNRAS.370..681N,2017MNRAS.471.2778B} and references therein) that the velocity distribution of elliptical galaxies and star clusters tends to be radially anisotropic. Moreover, non-spherical equilibrium systems (and elliptical galaxies might deviate considerably from spherical symmetry) have by construction anisotropic distribution functions, with different velocity dispersion profiles along their semi-axes (\citealt{1977MNRAS.181..735B}).\\  
\indent Surprisingly, relatively few works have analyzed DF in anisotropic systems, most of which concentrating on the effect of $\xi$ on the satellite orbit circularization (\citealt{1987IAUS..127..475C,2000ApJ...532..294T,2007A&A...463..921A,2017MNRAS.467..661N}). More recently, \cite{2025PhRvD.111f3071K}, investigating the DF acting on a a black hole moving on nearly circular orbits in a dark-matter cusp with central density proportional to $r^{-7/3}$ observed that increasingly higher values of the anisotropy parameter $\xi$ (i.e.\ the model is more and more dominated by low angular momentum orbits) result in a systematically less efficient DF with respect to the isotropic model with $\xi=1$.
Here, to explore whether the effect of the satellite size on the DF depends on the velocity distribution of the host, we performed, for different values of $r_{\rm sat}/r_{\rm gal}$, additional reduced $N$-body simulations in which the anisotropy index of the host was fixed to $\xi=1.7$, corresponding to the bona fide maximum value for stability in a Hernquist model (e.g. see \citealt{2002MNRAS.332..901N}). We note that in the adopted reduced $N$-body simulations, the radial orbit instability is artificially suppressed, because the host particles do not interact among themselves: however, we do not expect this instability to occur for the explored value of $\xi$.\\
\indent We found that in the explored radially anisotropic systems the efficiency of DF steadily decreases for increasing $r_{\rm sat}/r_{\rm gal}$, regardless of the specific mass ratio $\alpha$, as found for isotropic hosts. As an example, in Figure \ref{fig_aniso} we show the orbital decay of satellites with mass $M_{\rm sat}=10^{-2}M_{\rm gal}$ and scale radii $r_{\rm sat}=10^{-2}$ and $5\times 10^{-2}r_{\rm gal}$, initially placed on orbits with $e_0=0.9$, 0.24, 0.13 and 0.04 in a $\xi=1.7$ anisotropic Hernquist host. Consistently with what observed for the isotropic cases (marked with the thin dotted lines), the larger $r_{\rm sat}$ the longer is the decay time. We observe that, at least for the cases with lower values of $e_0$, a tendency of the DF of being slightly less effective in the anisotropic model. Moreover, we note that final displacement of the satellite from the centre of the host does not depend significantly on either $r_{\rm sat}$ or $\xi$ for all the explored values of $e_0$, thus implying a satellite mass- and/or resolution-driven origin.
\subsection{Effect of the satellite truncation radius $r_*$}\label{sec:rt}
\indent Tidal stripping and deformation alter significantly the efficiency of DF. We performed extra realizations of the full $N-$body experiments where $r_{\rm sat}$ is kept fixed and the truncation radius $r_*$ (which in Fig. \ref{fig_nbody} is $r_*=3r_{\rm sat}$) is varied between $2r_{\rm sat}$ and $8r_{\rm sat}$. We note in these experiments that $M_{\rm sat}$ and $r_{\rm sat}$ are kept fixed, so models with larger  $r_*/r_{\rm sat}$ have lower density profile. In the upper panel of Figure \ref{fig_cutoff} we show the evolution of $d$ for the case of a heavy satellite with $M_{\rm sat}=10^{-2}M_{\rm gal}$ and $r_{\rm sat}=0.1r_{\rm gal}$ on a moderately eccentric orbit with $e_0=0.24$. The DF is systematically less effective for increasing values of the truncation radius $r_*$. Consistently, we observe a significantly more important tidal mass stripping with increasing $r_*$, up until at least $t\approx 400t_c$, as shown in the right panel of Figure \ref{fig_cutoff}, where we plot as a function of time the mass fraction enclosed within a radius $r_*$ centered on the density peak of the satellite. For $t\gtrsim 500t_c$ with the only exception of the model with $r_*=2r_{\rm sat}$, the retained mass fraction relaxes on a value between roughly $1/2$ and $1/3$. Interestingly, the case with the weakest DF, corresponding to  satellite's initial conditions truncated at $r_*=8r_{\rm sat}$ is not the one that loses the largest fraction of its initial mass.

In Figure \ref{fig_tides} we report the particle positions in the orbital plane at $t/t_c=100$, 200 and 400  for the same simulations as in Fig.\ \ref{fig_cutoff} with $r_*/r_{\rm sat}=2$, 3, 4 and 8. As a general trend, increasing the truncation radius $r_*$ up to values larger that the initial $r_T$ implies a faster growth of the tidal arms and consequently a more efficient mass loss and eventually tidal disruption (\citealt{2003ApJ...596..314M,2004MNRAS.351..891Z,2007MNRAS.375..604F}). To isolate the effect of SF, we evolved the particle distribution representing the satellite in the smooth host potential (\ref{eq:hernquistphi}) without adding the extra DF term. In Figure \ref{fig_self} we show two sample cases with truncation radii $r_*=3r_{\rm sat}$ and $8r_{\rm sat}$. As a reference we also computed the unperturbed orbit, represented in the same figure by the black solid line. It appears that, over a time scale of about 15 pericentric passages of the reference orbit, as a consequence of the SF the more extended system (indicated by the empty red circles) experiences a stronger decay of the orbital apocentric distance $r_{\rm apo}$ (roughly the 10\%), while in both cases the pericentric distance $r_{\rm per}$ suffers little changes. For satellites placed on orbits with lower values of $e$, the effect of SF is, as expected, even lower with a reduction of about the 2\% in $r_{\rm apo}$ after 20 pericentric passages. 
\section{Relation to previous work and astrophysical implications}\label{sec:implication}
We have investigated via different schemes of $N-$body simulations the DF exerted by a spherical host system modelled with a Hernquist density profile on a satellite of finite extent $r_s$ initially placed on orbits of different eccentricity and specific energy. From the comparison of full self-consistent simulations (in which both the galaxy and the satellite are represented with particles) and simpler experiments (in which the satellite is represented by a single extended particle) with semi-analytical one particle integrations, we confirm previous indications (e.g. see \citealt{2006MNRAS.368.1073G,2010ApJ...725.1707G,2006MNRAS.373.1451R}) that DF is always significantly overestimated by the standard Chandrasekhar treatment.
We found that the finite extent of the satellite system (even in the idealized undeformable limit), considerably alters the effect of DF, with the least concentrated satellites (in units of the host typical density) being almost unaffected by the drag. In other words, using a standard point particle approximation to evaluate DF might grossly overestimate its time scale, in some cases even up to an order of magnitude for satellites on highly eccentric orbits with mass ratios of the order $\alpha\approx 10^{-3}$. We showed that even a correction of $\log\Lambda$ in the same fashion as \cite{2015MNRAS.454.3778P} remains model dependent and still overestimates the effect of the Chandrasekhar drag. As long as the mass loss is negligible (or, as evidenced by \cite{2007MNRAS.375..604F}, roughly the $50\%$ of the stripped mass remains sufficiently close to the remnant bound mass), \cite{1983A&A...117....9M} treatment of DF on an extended satellite is, among  the proposed semi-analytical models, the one that deviates less from the $N$-body results. 

The use of reduced $N$-body simulations allows one to disentangle the effect of a finite extent from the contribution of tidal deformation in slowing down the satellite. \cite{2020MNRAS.495.4496M}, who used a smooth potential to simulate the host and an $N$-body model for the satellite, estimated that the SF adds a significant contribution to DF only for the highly eccentric orbits. However, in our simulations we observe that this additional drag effect due to orbital angular momentum loss mediated by the tidal tails is almost entirely countered off by the mass loss itself. This is consistent, for example, with the $N$-body simulations of the Fornax dwarf spheroidal discussed by \cite{2024A&A...690A..61D,2025arXiv250803286D}, in which the DF and the tidal forces exerted by the Milky Way do not induce  a significant decay of Fornax pericentric distance.
\section{Conclusions}\label{sec:summary}
We studied the DF of a spherical host system on a satellite accounting for its finite size. We employed both full $N$-body simulations, both the host and the satellite are modelled with particles, and simplified $N$-body simulations in which only the host is represented with particles, either interacting (semi-reduced simulations) or non-interacting (reduced simulations) with themselves. 
We provided evidence that, though with some limitations, the very efficient reduced $N-$body simulations can be used to quantitatively investigate the effect of DF on realistic satellites. We quantified the dependence of the DF friction force on the size of the satellite: for both isotropic and radially anisotropic host, the DF timescales increases for increasing  satellite size, if the other parameters are kept fixed. We showed that all the proposed semi-analytical estimates of dynamical friction accounting for the satellite size via some parametrization of the Coulomb logarithm $\log\Lambda$ overestimate the speed of the orbital decay. Our full $N-$body simulations, allowing for mass loss and tidal deformation, indicate, consistently with \cite{2020MNRAS.495.4496M}, that the SF effect is rather mild, as confirmed also by ad hoc numerical experiments where the host is represented by a smooth potential and no DF can take place.
\begin{acknowledgements}
We thank Luca Ciotti, Pavel Kroupa, Bruno Marcos and Jorge Pe\~narrubia for useful discussions at an early stage of this project. One of us (PFDC) acknowledges support from the MUR PRIN2022 project “Breakdown of
ergodicity in classical and quantum many-body systems”
(BECQuMB) Grant No.\ 20222BHC9Z. 
\end{acknowledgements}
\bibliographystyle{aa}
\bibliography{ras}
\end{document}